\documentclass[structabstract]{aa}
\usepackage{graphicx}
\usepackage{txfonts}
\usepackage{verbatim}
\usepackage{natbib}

\begin{document}

\title{The continuous period search method and its application to the young solar analogue \object{HD 116956}
\thanks{The analysed photometry and numerical results of the analysis are both published electronically at the CDS.}}
\author{J. Lehtinen \inst{1}
\and L. Jetsu \inst{1} 
\and T. Hackman \inst{1,} \inst{2}
\and P. Kajatkari \inst{1}
\and G.W. Henry \inst{3}}
\institute{Department of Physics, Gustaf H\"{a}llstr\"{o}min katu 2a (P.O. Box 64), FI-00014 University of Helsinki, Finland
\and Finnish Centre for Astronomy with ESO, University of Turku, 
V\"{a}is\"{a}l\"{a}ntie 20, 21500 Piikki\"{o}, Finland
\and Center of Excellence in Information Systems, 
Tennessee State University, 3500 John A. Merritt Blvd., Box 9501, Nashville, TN 37209, USA}
\date{Received / Accepted}

\abstract{}
{We formulate an improved time series analysis method for the analysis of photometry of active stars. This new Continuous Period Search (CPS) method is applied to 12 years of $V$ band photometry of the young solar analogue \object{HD 116956} (\object{NQ UMa}).}
{The new method is developed from the previous Three Stage Period Analysis (TSPA) method. Our improvements are the use of a sliding window in choosing the modelled datasets, a criterion applied to select the best model for each dataset and the computation of the time scale of change of the light curve. We test the performance of CPS with simulated and real data.}
{The CPS has a much improved time resolution which allows us to better investigate fast evolution of stellar light curves. We can also separate between the cases when the data is best described by periodic (i.e. rotational modulation of brightness) and aperiodic (e.g. constant brightness) models. We find, however, that the performance of the CPS has certain limitations. It does not determine the correct model complexity in all cases, especially when the underlying light curve is constant and the number of observations too small. Also the sensitivity in detecting two close light curve minima is limited and it has a certain amount of intrinsic instability in its period estimation. Using the CPS, we find persistent active longitudes in the star \object{HD 116956} and a ``flip-flop'' event that occurred during the year 1999. Assuming that the surface differential rotation of the star causes observable period variations in the stellar light curve, we determine the differential rotation coefficient to be $|k|>0.11$. The mean timescale of change of the light curve during the whole 12 year observing period was $\overline{T}_{\rm C}=44.1$ d, which is of the same order as the predicted convective turnover time of the star. We also investigate the presence of activity cycles on the star, but do not find any conclusive evidence supporting them.}
{}
\maketitle

\section{Introduction}
\label{intro}

One of the most common methods to search for periodicity in an astronomical time series is the Lomb-Scargle periodogram \citep{lomb1976least, scargle1982studies}. The method is, however, not suitable for analysing data where the periodic variations do not fit to a simple single harmonic model. A more general period analysis method, the Three Stage Period Analysis (TSPA), was formulated by \citet[hereafter Paper I]{jetsu1999three}. This method utilises a higher order harmonic model: a $K$th order Fourier series. It can thus model more accurately periodic data with a complex form. The main motivation in developing the TSPA was to formulate a suitable method for analysing photometry of active stars.

The TSPA method can still be improved. In this paper we discuss and implement three such improvements. First of all, TSPA is not flexible enough for studying fast evolution of the light curve, as it divides the photometry into separate datasets that do not overlap. A better time resolution can be achieved by choosing the datasets with a sliding window and allowing the adjacent datasets to overlap. This produces a sequence of analysis results akin to video view in contrast to the set of snapshots provided by the TSPA. 

The second improvement is to test several models of different order $K$, instead of just one, as possible descriptions of the data. The best model can then be selected with a suitable criterion. By including a constant brightness model into the set of tested possible models, we can investigate whether a periodic function is an appropriate description of the data at all. This is of particular interest when analysing low amplitude stellar light curves.

As the third improvement to the TSPA, we define the parameter $T_{\rm C}$ as the approximate time scale in which the shape of the analysed light curve undergoes significant changes. It characterises the time interval after which any particular light curve model does not adequately describe the subsequent light curves any more. This parameter is, however, only an approximation, as it depends also on the quality and quantity of the photometric data. Nevertheless, when analysing photometry of active stars, the parameter may tell us how fast the spot distribution on the visible stellar surface is changing.

We formulate the new Continuous Period Search (CPS) method in Sect. \ref{cps} by incorporating the above three improvements to the TSPA, and use \emph{simulated} data to test the performance of the CPS in Sect. \ref{test}. Finally, we apply the CPS to \emph{real} photometric observations of the spot activity on the surface of the nearby young solar analogue \object{HD 116956} (\object{NQ UMa}) in Sect. \ref{demo}. 

\object{HD 116956} has a spectral type of G9V, an effective temperature of $T_{\rm eff}=5170$ K, a projected rotation velocity of $v\sin{i}=5.6$ km s${}^{-1}$ and a lithium abundance of $\log{N(\rm Li)}=1.42 \pm 0.12$ \citep{gaidos1998nearby}. From its space motion, \citet{gaidos2000spectroscopy} identified the star as a Local Association member. This would suggest an age between 20 and 150 Myr \citep{montes2001late}. On the other hand, \citet{gaidos2000spectroscopy} noted that none of the Local Association stars in their sample display characteristics of such a young age and rather suggest that they have similar ages to the Ursa Major group stars (300 Myr). Nevertheless, \object{HD 116956} must be considerably younger than the Sun, as also indicated by its lithium content. The radial velocity of \object{HD 116956} is constant $-12.3$ km s${}^{-1}$, i.e. it seems to be a single star \citep{gaidos2000spectroscopy}.

HD 116956 displays numerous signs of magnetic activity, such as rotational modulation of its brightness caused by large spots on its surface. The amplitude of this variability is typically 0.05 mag. \citet{gaidos2000spectroscopy} reported a preliminary rotation period estimate of $P=7.80\pm0.02$~days. Another activity indicator is the logarithmic ratio of X-ray and bolometric luminosities $R_{\rm X}=-4.48$, which indicates that the star has an active corona \citep{gaidos1998nearby}. Finally, the logarithmic ratio of  Ca II emission and bolometric luminosity, $R'_{\rm HK}=-4.447$, means that \object{HD 116956} is chromospherically active \citep{gray2003contributions}.

\section{The CPS method}
\label{cps}

The light curves of active stars can undergo rapid changes, but may as well remain stable over several rotations. These light curves have usually been modelled with one simple sinusoid for the whole data or for fixed parts of it. In reality, the time span of observations required for reliable light curve modelling cannot be determined uniquely, because it is certainly not constant. Nor does the best model for the light curve remain the same, e.g. periodicity can vanish and another periodicity can reappear at some later epoch. This means that the selection of the slices of data, hereafter called datasets, must be flexible and the complexity of the model must adapt to the light curve changes. The CPS method utilises overlapping datasets, identifies the best model for each of these and gives a quantitative estimate for the temporal stability of these models. Although the CPS is applied here only to simulated and real ground based photometry of active stars, this method can be easily adjusted to search for periodicity in any type of unevenly spaced data.

\subsection{Datasets and segments}
\label{set}

The input data for the CPS method are observations $y_i=y(t_i)$ at time points $t_i$ with observational errors $\sigma_i$. Before any modelling is done, the data must be divided into short datasets (SET), each of which is modelled individually. The length of these datasets is $\Delta T=t_n-t_1$, which we limit to have the maximum length of

\begin{eqnarray}
\Delta T_{\rm max} \!=\!
\left\{
{\begin{array}{rl}
\Delta T_1 \ , & P_0 < \frac{1}{2}\Delta T_1 \\
2P_0 \ , & \frac{1}{2}\Delta T_1 \leq P_0 \leq \frac{1}{2}\Delta T_2 \\
\Delta T_2 \ , & P_0 > \frac{1}{2}\Delta T_2,
\end{array}}
\right.
\label{deltatmax}
\end{eqnarray}

\noindent where $P_0$ is the a priori estimate for the photometric rotation period. The lower limit $\Delta T_1$ ensures that most of the datasets contain enough observations for modelling. Datasets with $n<10$ are not analysed with the CPS. The upper limit $\Delta T_2$ is necessary to prevent significant changes in the light curve during a single dataset for stars with longer rotation periods. In other words, the parameter $\Delta T_{\rm max}$ is applied to limit the length $\Delta T=t_n-t_1$ of the analysed datasets so that they yield a consistent light curve model but still have enough observations for reliable modelling. In practice, suitable values for $\Delta T_1$ and $\Delta T_2$ are 25 d and 200 d respectively, when analysing ground based photometry of active stars. For different kinds of data, the values of $\Delta T_1$ and $\Delta T_2$ must be determined separately.

Unlike in the TSPA, the datasets are allowed to overlap in the CPS. This gives a better time resolution. In typical ground based observations, we begin a new dataset at the first time point $t_1$ of each night and include all data points $y_i=y(t_i)$ within $t_1 \leq t_i \leq t_1+\Delta T_{\rm max}$. A candidate for the next consecutive dataset is chosen by moving the window one night forward. We require that the observations of two consecutive datasets fulfill the condition

\begin{equation}
{\rm SET}_k \not\subset {\rm SET}_{k+1} \ \mathrm{and} \ {\rm SET}_{k+1} \not\subset {\rm SET}_k.
\end{equation}

\noindent In other words, both datasets must contain at least one observation that does not belong to the other one. If this condition is violated, the candidate dataset $\rm SET_{k+1}$ is rejected. In this case, the dataset window is moved one night further. If there is much data during each night and $P_0$ is short, a denser selection of datasets might be desirable. In this case, one may even start a new dataset at each new observation. This kind of a much denser dataset selection rule would be ideal for satellite photometry.

Observations that would be included only in datasets having $n < 10$ are of little use for modelling with the CPS, as for them $n$ approaches the number of free parameters of the model. Such observations are rejected as temporally isolated data points.

The data are further divided into longer segments (SEG) whenever there is a gap longer than $\Delta T_{\rm max}$ in the observations. Such gaps may have contained rejected isolated data points. These segments are defined solely for the purpose of identifying datasets belonging to different observing seasons.

The benefit of letting the subsets overlap is a better time resolution compared to the TSPA, where the datasets were separated from each other. As a drawback, the modelling results from two overlapping datasets are correlated with each other. One can easily eliminate this correlation by comparing only the non-overlapping datasets. The alternatives in selecting these non-overlapping (hereafter called the \emph{independent}) datasets are numerous. The most simple alternative is to select the independent datasets by beginning from the first dataset of each segment and then selecting the next independent dataset with the criterion that it does not have any common data points with the preceding independent dataset.

\subsection{Modelling of the observations}
\label{modelling}

The TSPA model

\begin{equation}
\hat{y}(t_i) = \hat{y}(t_i,\bar{\beta}) = M + \sum_{k=1}^K{[B_k\cos{(k2\pi ft_i)} + C_k\sin{(k2\pi ft_i)}]},
\label{model}
\end{equation}

\noindent i.e. a $K$th order Fourier series, is also used in the CPS. This model has an order $K$ and $2K+2$ free parameters, $\bar{\beta}=[M,B_1,\ldots,B_K,C_1,\ldots,C_K,f]$. In other words, the free parameters are the mean $M$, the individual cosine and sine amplitudes $B_k$ and $C_k$ and the frequency $f$.

In the TSPA, the modelling proceeded through three stages: the pilot search, the grid search and the refined search. In the CPS, the pilot search, where the initial period estimates were identified within a broad frequency interval (Paper I, Sect. 3.1), is not used. Rather, the grid search (Paper I, Sect. 3.2) is performed straight away. In the grid search a dense grid of frequencies is tested one at the time by fitting the model (Eq. \ref{model}) to the observations. When the frequency $f$ is fixed, the model becomes linear and the remaining free parameters $[M,B_1,\ldots,B_K,C_1,\ldots,C_K]$ of the model have unique solutions. The grid search is performed within the period range of

\begin{equation}
(1-q)P_0 \leq P \leq (1+q)P_0,
\label{gridpp}
\end{equation}

\noindent where $P_0$ is the a priori rotation period estimate of Eq. \ref{deltatmax} and $q$ regulates the interval of the period search. We use a value of $q=0.15$. This is usually sufficient, since for many active stars there already exist period determinations in the literature and the real period changes in photometry can be expected to fall inside the $\pm15\%$ range of $P_0$. If necessary, this tested period range can be expanded. If no $P_0$ is available, one can be obtained, for example, with the TSPA.

The final model parameters $\bar{\beta}$ are obtained from the refined search (Paper I, Sect. 3.3), which consists of a standard nonlinear Marquardt iteration that minimises $\chi^2$ with weights,

\begin{equation}
\chi^2(\bar{y},\bar{\beta}) = \sum_{i=1}^n{w_i\epsilon_i^2},
\label{chi2}
\end{equation}

\noindent where $w_i=\sigma_i^{-2}$ are the weights and $\epsilon_i=y_i-\hat{y}(t_i,\bar{\beta})$ are the residuals. The refined search can find the correct period even if it is outside but close to the grid search range (Eq. \ref{gridpp}).

Before modelling, the data is preprocessed by removing outliers. These outliers are determined using a preliminary model of order $K'$. We set this preliminary modelling order equal to the highest model order used in the actual modelling of the data, $K'=K_{\rm lim}$.  The preliminary model $\hat{y}'(t)$ gives the residuals $\epsilon_i' = y_i-\hat{y}'(t_i)$. For each dataset, the observations having residuals larger than three times the standard deviation of all residuals $\bar{\epsilon}'$ are removed as outliers. For our data, we found that there are typically only a few such outliers among several hundreds of observations.

The problem of determining the order $K$ of the Fourier model (Eq. \ref{model}) is crucial to the light curve modelling procedure. The value of $K$ must be high enough to allow a good fit to the observed light curve but not too high to result in overfitting to the data. In the TSPA, the order $K$ was selected beforehand and the same value was used for all datasets. In contrast, the CPS uses a Bayesian information criterion to select the best $K$ value separately for each dataset. This criterion consists of minimising a type of $\chi^2$ with an additional penalty term for the extra degrees of freedom introduced by a higher order $K$.

Before testing for the optimal $K$ for each dataset, the upper limit $K_{\rm lim}$ for the highest accepted modelling order must be selected. Since the smooth light curves of spotted stars usually display only one or two minima, not very high values of $K_{\rm lim}$ are necessary.  For more complicated light curve shapes, such as those of eclipsing binaries, higher $K_{\rm lim}$ values should be used to achieve a satisfactory fit. In practice, the value of $K_{\rm lim}=2$ is sufficient for analysing ground based photometry of most spotted stars.

The problem of determining $K$ is equivalent to finding the number of sinusoids embedded in the data. This number is just the order $K$ of the model. A solution for this particular problem was presented by \citet{stoica2004model}, who studied the applicability of different model order selection rules. They argued that the best performance is achieved with the Bayesian information criterion. It has the property that it always finds the correct $K$ as the number of data points $n$ increases to infinity.

The Bayesian information criterion is given by

\begin{equation}
R_{\rm BIC} = 2n\ln{\lambda(\bar{y},\bar{\beta})} + (5K+1)\ln{n},
\label{bic}
\end{equation}

\noindent where the first term gives the logarithmic likelihood with $\lambda(\bar{y},\bar{\beta}) = \chi^2(\bar{y},\bar{\beta})\left[\sum_{i=1}^n{w_i}\right]^{-1}$. The criterion is used by first modelling the dataset separately with all models between $K=0$ and $K=K_{\rm lim}$ and then calculating $R_{\rm BIC}$ for each of these models. The model with $K=0$ (i.e. constant) is obtained simply from the weighted mean $m_y$ of all data points $y_i$ in the dataset, i.e. $\hat{y}(t_i)=M=m_y$. The model for the dataset that minimises $R_{\rm BIC}$ has the optimal order $K$ and is chosen as the best model for the dataset.

The function of the two terms in Eq. \ref{bic} are the following. The first term describes the goodness of the fit to the data. It is naturally smaller for higher order models, as they allow a more detailed fit. The second term, $(5K+1)\ln{n}$, is a penalty term which grows linearly as a function of $K$. It balances the smaller values of of the first term $2n\ln{\lambda(\bar{y},\bar{\beta})}$ at a high $K$ and thus prevents overfitting. Because of this second term there will be an optimal $K$ where $R_{\rm BIC}$ reaches its minimum value and then starts to grow again as $K$ increases.

Most of the free parameters of the model, $\bar{\beta} = [M,B_1,\ldots,B_K,C_1,\ldots,C_K,f]$, are not very useful as such. The physically meaningful parameters are the mean magnitude $M$ and the total amplitude $A$ of the light curve, the period $P=f^{-1}$, as well as the epochs $t_{\rm min,1}$ and $t_{\rm min,2}$ of the primary and secondary minima of the light curve. The mean $M$ and the period $P=f^{-1}$ are obtained directly from the free parameters $\bar{\beta}$ of the model, but the values of $A$, $t_{\rm min,1}$ and $t_{\rm min,2}$ must be determined numerically from the light curve model. Note that only the models with $K\geq2$ can provide all of these parameters. For $K=1$, the secondary minimum $t_{\rm min,2}$ does not exist and the only parameter that can be obtained for $K=0$ is the mean $M=m_y$.

The error estimates and reliabilities of these parameters are determined as in the TSPA (Paper I, Sects. 4 and 6.3). This process consists of running a bootstrap with 200 rounds for the residuals $\bar{\epsilon}$ and all model parameters. The errors of the parameters are the standard deviations of the 200 bootstrap estimates. All these bootstrap distributions are tested against the Gaussian hypothesis

\begin{itemize}
\item[H$_{\rm G}$:] \emph{$\bar{x}$ represents a random sample drawn from a Gaussian distribution},
\end{itemize}

\noindent where $\bar{x}$ denotes the residuals or the bootstrap distributions of any of the model parameters. If any of these distributions fails to satisfy H$_{\rm G}$, all of the model parameters are considered unreliable. The Gaussian hypothesis H$_{\rm G}$ is tested using the Kolmogorov-Smirnov test with preassigned significance level $\gamma=0.01$ for rejection. Also, if the secondary minimum $t_{\rm min,2}$ is present in less than 95\% of the bootstrap samples, it is considered unreliable.

\subsection{Time scale of change}
\label{correlation}

The modelling gives the mean $M(\tau)$, total amplitude $A(\tau)$, period $P(\tau)$ and the minimum epochs $t_{\rm min,1}(\tau)$ and $t_{\rm min,2}(\tau)$ of the light curve, where $\tau$ is the mean of all observing times $t_i$ in the current dataset. As the shape of the light curve usually evolves with time, the model must also evolve. It is of interest to investigate for how long an unchanged model for a specific dataset still reasonably well fits with the observations of the subsequent datasets. We estimate the time scale of change $T_{\rm C}$ for each dataset $\iota$ by determining for how many of the following datasets $\iota+\kappa$ the model of dataset $\iota$ still fits the data.

The dataset $\iota$ contains the observations $\bar{y}_{\iota}$ which yield a model $\hat{y}_{\iota}(\bar{t}_{\iota})$. Another dataset $\iota+\kappa$ at a later epoch contains the data $\bar{y}_{\iota+\kappa}$. Assuming that the model $\hat{y}_{\iota}(\bar{t}_{\iota})$ is valid for both datasets, we define the residuals

\begin{equation}
\bar{\epsilon}_{\iota} = \bar{y}_{\iota} - \hat{y}_{\iota}(\bar{t}_{\iota})
\end{equation}

\noindent and

\begin{equation}
\bar{\epsilon}_{\iota,\kappa} = \bar{y}_{\iota+\kappa} - \hat{y}_{\iota}(\bar{t}_{\iota+\kappa}).
\end{equation}

If the light curve has not changed between datasets $\iota$ and $\iota+\kappa$, the residuals $\bar{\epsilon}_{\iota}$ and $\bar{\epsilon}_{\iota,\kappa}$ should have similar distributions, i.e. they should satisfy the hypothesis

\begin{itemize}
\item[H$_{\rm K2}$:] \emph{$\bar{\epsilon}_{\iota,\kappa}$ and $\bar{\epsilon}_\iota$ represent random samples drawn from the same distribution}.
\end{itemize}

\noindent H$_{\rm K2}$ is tested using the two-sided Kolmogorov-Smirnov test. The preassigned significance level for rejecting H$_{\rm K2}$ is $\gamma=0.01$. The time scale of change is not determined for datasets where the model parameters have been found unreliable. This means that we already know that the residuals $\bar{\epsilon}_\iota$ follow a Gaussian distribution. Thus the distribution of $\bar{\epsilon}_{\iota,\kappa}$ should also resemble a Gaussian distribution, if H$_{\rm K2}$ is not rejected.

The computation of the time scale of change for the model of dataset $\iota$ starts from the dataset $\iota+\kappa$, where $\kappa=1$. If the sets of residuals $\bar{\epsilon}_{\iota}$ and $\bar{\epsilon}_{\iota,\kappa=1}$ for the two datasets pass the test for H$_{\rm K2}$, the comparison of residuals proceeds to the next dataset, i.e. $\kappa\rightarrow\kappa+1$, and the same test is applied again. Finally, when this test fails, the comparison process is terminated. In this case, the time difference between the dataset $\iota$ and the last dataset passing the test, $\iota-\kappa-1$, is taken as the time scale of change for the subset $\iota$, i.e.

\begin{equation}
T_{\rm C}(\tau_{\iota})=\tau_{\iota+\kappa-1}-\tau_{\iota}.
\label{tc}
\end{equation}

\noindent
This computation of $T_{\rm C}$ only proceeds until the end of each segment. No datasets from other segments are compared, because there is no data confirming or refuting the model during the gap between the two segments and this might introduce a bias into the $T_{\rm C}$ estimates. If the computation process hits the end of the segment, a special value $T_{\rm C}=-2$ is given for the correlation time scale. This denotes the case that this particular model describes adequately all the datasets following it within the segment.

\section{Testing the method}
\label{test}

Before analysing real stellar photometry, we used simulated data to determine the performance of the CPS method in certain critical situations. This allowed us to get more insight in understanding some of the results of the analysis. Failing to do this can lead to a biased interpretation of the results.

Since the distribution of the observing times can have a profound effect to the results of the analysis, we generated our simulated test data using real observing times of the star HD 116956. Thus, any spurious effects introduced by the observing times alone should also be present in the analysis of both the simulated data and the real data of HD 116956 (Sect. \ref{demo}). Because other active stars are observed in a similar manner, the results of this section are applicable also when analysing them.

The observing times $t_i$ used for generating the simulated test data were taken from the first 152 observations of the time series of HD 116956. These form a complete unbroken observing season and are temporally fairly evenly distributed. To get a longer time series for the simulations, we appended a duplicated set of $t_i+t_{152}-t_1$ to the original set of $t_i$ and removed the double occurrence of $t_{152}$. This resulted in a set of 303 observation times with a complete time span of 377.43 d.

When simulating the periodic test data, we used a period of $P_0=7.8$ d. This imitated the a priori period estimate for HD 116956. With the chosen observing times and this particular period, our test analyses each yielded a total of 184 datasets with $\Delta T_{\rm max}=25$ d, where the number of simulated observations in each dataset was in the range $12 \leq n \leq 28$.

\subsection{Sensitivity to changes in the model order}

As already stated in Sect. \ref{modelling}, the Bayesian information criterion (Eq. \ref{bic}) should determine the correct modelling order with a very high probability when the number of observations is large. However, when analysing photometry of active stars, this number of observations is often quite limited. Typically we have $10 \leq n \leq 20$ in a dataset. This scarcity of data can potentially affect the performance of the model order selection.

To test the performance of the model order selection, we analysed three different simulated time series. These consisted of one set of constant data and two sets of sinusoidal data each with random Gaussian noise

\begin{eqnarray}
{\begin{array}{ll}
y_{K=0}(t_i) = \epsilon_{{\rm N},i} \\
y_{K=1}(t_i) = 0.01\sin{(2\pi ft_i)} + \epsilon_{{\rm N},i} \\
y_{K=2}(t_i) = 0.01\sin{(4\pi ft_i)} + \epsilon_{{\rm N},i}.
\end{array}}
\label{genmod}
\end{eqnarray}

\begin{figure}
\resizebox{\hsize}{!}{\includegraphics[angle=90]{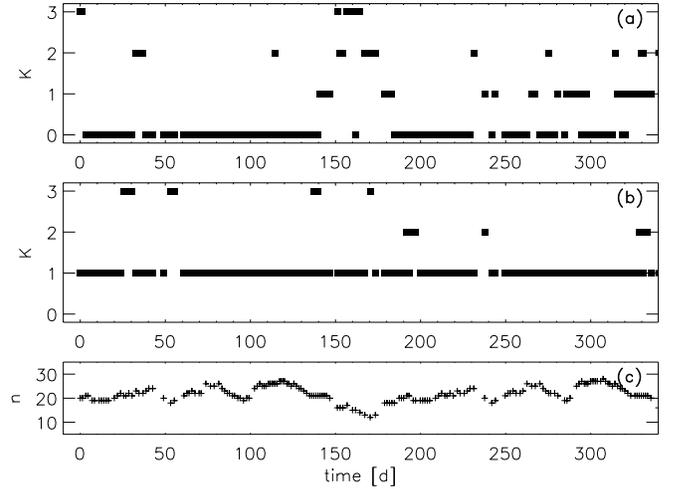}}
\caption{Model order $K$ from analysis of test data simulated with (a) model $y_{K=0}(t_i)$ in Eq. \ref{genmod} and (b) model $y_{K=1}(t_i)$  in Eq. \ref{genmod} having amplitude to noise ratio $\rm A/N=10$. The number of data points in each dataset is presented in panel (c).}
\label{kpic}
\end{figure}

\begin{table}[t]
\caption{Fractions of falsely determined model order $K$  
in the analysis stable simulated data.}
\center
\addtolength{\tabcolsep}{-0.08cm}   
\begin{tabular}{c c c c c c c}
\hline\hline
$n$ & $K$ & $\rm A/N=0$ & $\rm A/N=2$ & $\rm A/N=5$ & $\rm A/N=10$ & $\rm A/N=20$ \\
\hline
$n\leq20$ & 0 & 0.39 & \ldots & \ldots & \ldots & \ldots \\
& 1 & \ldots & 0.16 & 0.18 & 0.12 & 0.15 \\
& 2 & \ldots & 0.17 & 0.15 & 0.13 & 0.16 \\
\hline
$n>20$ & 0 & 0.21 & \ldots & \ldots & \ldots & \ldots \\
& 1 & \ldots & 0.10 & 0.10 & 0.06 & 0.07 \\
& 2 & \ldots & 0.08 & 0.09 & 0.07 & 0.10 \\
\hline
\end{tabular}
\addtolength{\tabcolsep}{+0.08cm}   
\label{ktab}
\end{table}

\noindent Here $f^{-1}=P_0=7.8$ d and $\epsilon_{{\rm N},i}$ denotes Gaussian noise distribution having a zero mean and a variance $\sigma_{\rm N}^2$, i.e. $\epsilon_{{\rm N},i} \sim N(0,\sigma_{\rm N}^2)$. These three simulation models were chosen so that the correct modelling order should be $K=0$, $K=1$ and $K=2$, respectively. Although both $y_{K=1}(t_i)$ and $y_{K=2}(t_i)$ are simple sinusoids, the CPS should naturally select $K=2$ to model the data generated with $y_{K=2}(t_i)$ when all tested period candidates are close to $P_0$. Then the CPS has to use a higher order model to correctly describe the two minima of $y_{K=2}(t_i)$ occurring with the period $P_0$. All the analyses were performed with $K_{\rm lim}=3$. This allowed us to examine the possibility that the criterion of Eq. \ref{bic} may yield a too high $K$ even when analysing simulated data generated with $y_{K=2}(t_i)$.

In the cases of $y_{K=1}(t_i)$ and $y_{K=2}(t_i)$, the test data were simulated using four different noise levels where $\sigma_{\rm N}$ was  0.005, 0.002, 0.001 or 0.0005. Given the light curve half amplitude $A/2=0.01$, these correspond to amplitude to noise ratios $\rm A/N=2$, $\rm A/N=5$, $\rm A/N=10$ and $\rm A/N=20$, using the definition ${\rm A/N}=A/2\sigma_{\rm N}$. According to this definition, the data simulated with $y_{K=0}(t_i)$ and any $\sigma_{\rm N}$ has $A/N=0$, and thus the results for $K=0$ are independent of the absolute level of noise.

After analysing the simulated test data, we checked how often the CPS arrived at a $K$ value different from that used in the simulation model. Fractions of false $K$ in 15 simulated time series (a total of 2760 analysed datasets) are presented in Table \ref{ktab}. They are reported separately for those datasets that had $n\leq20$ and those that had $n>20$. In the case of $y_{K=1}(t_i)$ and $y_{K=2}(t_i)$, the probability for a false model selection seems to be quite insensitive to $\rm A/N$. There is no clear structure visible in the failure probabilities at different $\rm A/N$, nor between $K=1$ and $K=2$. This seems to indicate that the sensitivity of the selection criterion does not strongly depend on $\rm A/N$ or the complexity of the periodic model. On the other hand, the number of observations in the datasets has a substantial influence to its success rate so that with more observations the correct model order is detected with a much higher probability. Nevertheless, the success rate stays between 80\% and 95\% in both the cases $n\leq20$ and $n>20$, which can be considered as a satisfactory performance.

Unfortunately, in the case of $y_{K=0}(t_i)$, i.e. pure noise, the CPS still detected spurious periodicity in 21\% of the datasets having $n>20$ and 39\% of those having $n\leq 20$. This can cause problems if the analysis results are not interpreted critically. Luckily, these spurious periodic models can often be identified from their very low amplitudes. As a rule, the amplitudes of these models have values comparable to the preset noise level $\sigma_{\rm N}$ and the observed standard deviation of the residuals $\sigma_{\epsilon}$ so that they fulfill $A \approx 2\sigma_{\rm N} \approx 2\sigma_{\epsilon}$. This is expected, as the amplitudes are artifacts resulting solely from the modelling of noise.

\begin{figure}
\resizebox{\hsize}{!}{\includegraphics[angle=90]{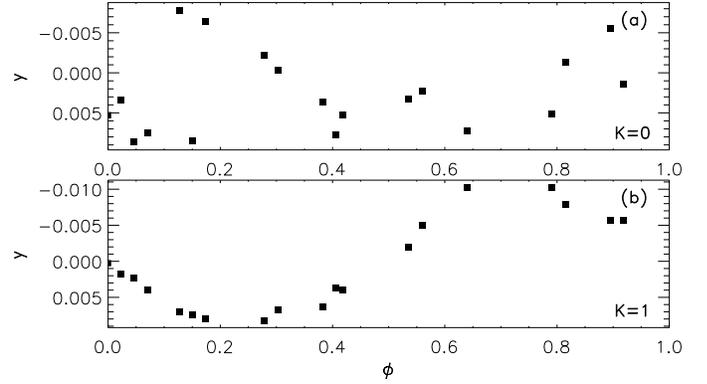}}
\caption{Examples of single 25 d long datasets simulated with (a) model $y_{K=0}(t_i)$ in Eq. \ref{genmod} and (b) model $y_{K=1}(t_i)$  in Eq. \ref{genmod} having amplitude to noise ratio $\rm A/N=10$. The phases of these simulated observations are calculated using the period of the simulation model, $P_0=7.8$ d.}
\label{kex}
\end{figure}

Examples of the behaviour of $K$ during a test analysis for typical simulated data are presented graphically in Fig. \ref{kpic}. The analysis of pure noise displays a large scatter of false results, although the correct value $K=0$ is still clearly the most commonly obtained result (Fig. \ref{kpic} (a)). The scatter is large especially near the mid section of the time series, where the number of simulated observations is small, $n<20$. The analysis of the data simulated with $y_{K=1}(t_i)$ and $\rm A/N=10$ behaves much better with only some overfitting and no false detections of $K=0$ models (Fig. \ref{kpic} (b)).

Clues to why the Bayesian information criterion has trouble in detecting the correct $K$ for the data simulated using the $y_{K=0}(t_i)$ model can be seen in Fig. \ref{kex}. This figure presents simulated observations during two arbitrary simulated datasets and folded into phase representation using the model period $P_0=7.8$ d. Because of the relatively small number of available data points ($n=19$), the case of pure noise, i.e. underlying $K=0$, still vaguely resembles some periodic form (Fig. \ref{kex} (a)). The probability of chance resemblance of periodic data decreases only for a larger number of data. In the case of an underlying $K=1$ simulation model (Fig. \ref{kex} (b)) it is quite clear that a single sinusoid is the correct model for the data.

We conclude that the CPS detects real periodicity even from low amplitude data of a high noise level. If there is no real periodicity, spurious periodicity is still detected in about a quarter of the analysed datasets. These spurious period detections can, however, be identified because their observed amplitude to noise ratio is near unity,

\begin{eqnarray*}
({\rm A/N})_{\rm obs} = \frac{A}{2\sigma_{\epsilon}} \approx 1.
\end{eqnarray*}

\noindent In conclusion, the results need always to be interpreted with special care when the dataset contains few observations and the model amplitude $A$ is low.

\subsection{Sensitivity to two close minima}
\label{testmin}

\begin{figure}
\resizebox{\hsize}{!}{\includegraphics[angle=90]{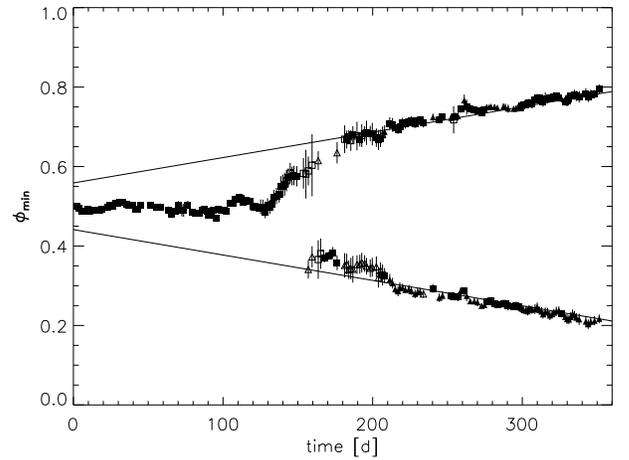}}
\caption{Primary (squares) and secondary (triangles) minimum phases from analysis of test data simulated with $y_{\rm spot}(t_i/P_0)$ of Eq. \ref{spotspot}. Filled symbols denote reliable and open symbols unreliable phase estimates. The two straight lines denote the correct phases where the simulated spots should have been detected.}
\label{genspot}
\end{figure}

When analysing the distribution of starspots on the surface of a star, one often simply identifies the light curve minima as direct markers of the spot longitudes. In the case of a single spot or two clearly separated ones, this assumption, of course, holds and the light curve minimum phases give good approximations of the spot longitudes.

However, if two spots are longitudinally close enough to each other, the observable light curve minima can be significantly shifted. As the two minima produced by the two spots move closer to each other, their tails start to merge. At first, the minima stay separated but are shifted somewhat towards each other. At phase separations below some critical value $\Delta \phi_{\rm crit}$, the minima finally merge and produce one single broad minimum. The observed phase of this single minimum lies between the phases of the two underlying spots.

The tendency of the light curve minima, produced by two individual spots, to merge depends on their width. Note, that no minimum can be very narrow, even in stars with an inclination close to $90^{\circ}$. For example, a spot located at the stellar equator is typically visible over half of the stellar rotation. During this time it modulates the visible brightness of the star. Thus the total width of the minimum spans about one half of the total rotation period. If the spot lies closer to the visible pole or the stellar inclination is small, it stays at the visible stellar disk for a longer period and causes an even broader minimum. In the extreme case, a spot can stay at the visible hemisphere at all times. In this case it causes a very broad minimum induced only by its variable projected area and limb darkening. Only a spot below the equator, closer to the unseen pole, may cause a narrow minimum, as it stays visible for less than half of the rotation period. However, all minima significantly narrower than half of the rotation can only be produced by spots that appear on the visible hemisphere briefly and never come very far from the limb of the star. Because of limb darkening and a small projected spot area, these minima are very shallow and are therefore strongly affected by noise.

\begin{figure}
\resizebox{\hsize}{!}{\includegraphics[angle=90]{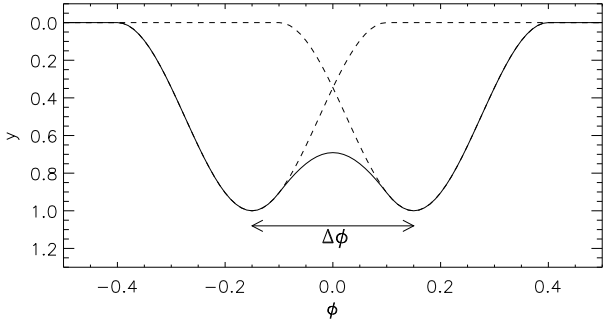}}
\caption{Two single spot model light curves according to $y_{\rm spot,1}(\phi)$ plotted around phases $\phi_0=0.15$ and $\phi_0=-0.15$ (dashed lines) and the combined light curve (solid line). The phase separation $\Delta\phi$ is indicated with the two headed arrow.}
\label{yspot}
\end{figure}

To get an estimate of the critical phase separation $\Delta \phi_{\rm crit}$, consider a model light curve of the form

\begin{eqnarray}
y_{\rm spot,1}(\phi) \!=\!
\left\{
{\begin{array}{ll}
0 \ , &  \phi < -0.25 \\
\frac{1}{2}[1 + \cos{(4\pi\phi)}] \ , & -0.25 \leq \phi \leq 0.25 \\
0 \ , & \phi > 0.25
\end{array}}
\right.
\label{spot1}
\end{eqnarray}

\noindent in phase space within the interval $-0.5 < \phi \leq 0.5$. This model approximates reasonably well the shape of a light curve produced by a single spot at the stellar equator at phase $\phi=0$. Putting two equally strong spots at phases $\phi_0$ and $-\phi_0$ results in a merged light curve. The merged light curve within the interval $\phi_0-0.25 \leq \phi \leq 0.25-\phi_0$ has the form

\begin{eqnarray*}
y_{\rm spot,2}(\phi) = \frac{1}{2}[2 + \cos{(4\pi(\phi+\phi_0))} + \cos{(4\pi(\phi-\phi_0))}].
\end{eqnarray*}

\noindent Note that this is only a linear approximation, as actual photometry is measured in magnitudes which can not be added in this manner themselves. In the case of starspot induced light variability where the amplitude of the light curve is in the range of $0.01-0.1$ mag, this linear approximation can, however, be done. The combination of two $y_{\rm spot,1}(\phi)$ curves at the phases $\phi_0=0.15$ and $\phi_0=-0.15$ is visualised in Fig. \ref{yspot}. The two underlying curves are drawn with dashed lines and the merged curve with a solid line. The phase separation of the two components is indicated with the two headed arrow.

The two spots can be identified from the light curve if they cause two separate light curve minima. In other words, there should be a secondary maximum at phase $\phi=0$ between the two minima. This corresponds to the condition

\begin{eqnarray*}
y_{\rm spot,2}''(0) = 
-8\pi^2[\cos{(4\pi(\phi_0))} + \cos{(4\pi(-\phi_0))}] &<& 0 \\
\Rightarrow \cos{(4\pi\phi_0)} &>& 0.
\end{eqnarray*}

\noindent For $0 \leq \phi_0 \leq 0.25$, this corresponds to the phase separation of the two spots being

\begin{equation}
\Delta\phi = 2\phi_0 > 0.25,
\label{criticallimit}
\end{equation}

\noindent i.e. no spots closer to each other than $\Delta\phi_{\rm crit}=0.25$ should cause two separate light minima. Less than optimal performance of the analysis algorithm may yield an even larger value of $\Delta\phi_{\rm crit}$.

To determine the phase resolution of the CPS, we tested a two spot simulation model based on two adaptions of Eq. \ref{spot1},

\begin{equation}
y_{\rm spot}\left(\frac{t_i}{P_0}\right) =
y_{\rm spot,1}\left(\frac{t_i}{P_0}\right) + 
y_{\rm spot,1}\left(\frac{t_i}{P_0+\Delta P}+0.1\right) + \epsilon_{{\rm N},i}.
\label{spotspot}
\end{equation}

\noindent The model describes two spots rotating with periods $P_0$ and $P_0+\Delta P$ and having an initial phase separation of $\Delta\phi=0.1$. The difference in period, $\Delta P=-0.001327P_0$, was chosen so that at the end of the 377.43 d long time series the phase difference between the two spots would have increased from $\Delta\phi=0.1$ to 0.6. The random noise used in simulating this time series was $\epsilon_{{\rm N},i} \sim N(0,0.001^2)$, i.e. $\rm A/N=10$.

The resulting minimum phases from the CPS analysis of this simulated data are shown in Fig. \ref{genspot}. Superimposed on the minimum phases detected with the CPS are two straight lines indicating the correct simulated phases where the two spots should have been found. At the beginning of the time series, the two underlying minima are inseparable and produce one common merged minimum with a phase between the two correct phases, as expected. Towards the end of the time series, the two minima both become detectable and have estimated phases very close to their correct values. However, near the time when the two minima first become separable their estimated phase differences are systematically smaller than the correct simulated values.

The two minima are first detected separately at time $t=167.99$ d at which time their phase separation given by the CPS is $\Delta\phi=0.27$. The simulated correct phase separation at that moment is $\Delta\phi=0.32$. Both of these two values exceed the limit $\Delta\phi_{\rm crit}=0.25$ derived in Eq. \ref{criticallimit}. This indicates that the ability of the CPS to separate individual minima is not perfect. We conclude that spots with a phase separation smaller than about one third of a rotation are not expected to be separable with the CPS.

The failure to detect two close by spots separately from a light curve means that any time when two spots are detected from a light curve, they already have a large longitudinal separation. In other words, our impression of the distribution of the spots is biased towards spots lying nearly at opposite sides of the star. This can lead to detections of spurious active longitudes even in the case when the underlying physical spot distribution has a completely different geometry.

\subsection{Stability of the method}
\label{teststab}

The results of the analysis should be reliable regardless of random effects caused by noise and uneven distribution of the observing times. Analysis of test data simulated with an unchanging underlying model should thus give stable parameter estimates throughout the whole time series.

\begin{figure}
\resizebox{\hsize}{!}{\includegraphics[angle=90]{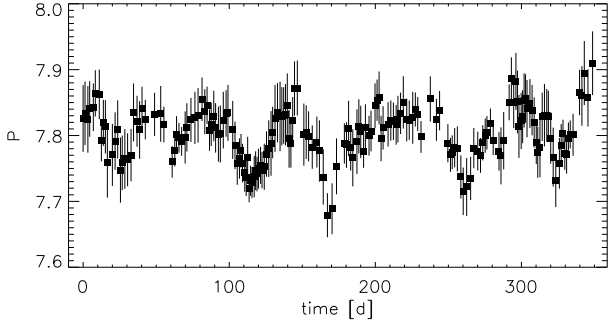}}
\caption{Period estimates from the analysis of test data simulated with $y_{K=1}(t_i)$ having $\rm A/N=10$ (see Eq. \ref{genmod}).}
\label{genp}
\end{figure}

To test the stability of the CPS, we analysed sinusoidal test data simulated with $y_{K=1}(t_i)$ (Eq. \ref{genmod}). The standard deviation of the noise was selected from $\sigma_{\rm N} \in [0.0005,\ 0.0007,\ 0.001,\ 0.002,\ 0.003,\ 0.005]$ corresponding to amplitude to noise ratios $\rm A/N=20$, $\rm A/N=14.3$, $\rm A/N=10$, $\rm A/N=5$, $\rm A/N=3.3$ and $\rm A/N=2$, respectively. The analysis yielded generally rather stable values for the simulated light curve mean $M$ and amplitude $A$ and the single minimum phase $\phi_{\rm min,1}$. However, the period estimate $P$ showed quite significant variations, as can be seen in Fig. \ref{genp}. This type of spurious variations must be taken into account, since the period changes of spotted stars are commonly used to measure the strength of their surface differential rotation \citep{jetsu1993decade}.

\begin{table}[t]
\caption{The spurious changes of the period $P$ in units of $Z$ (Eq. \ref{z}) for test data simulated with $y_{K=1}(t_i)$ (Eq. \ref{genmod}) and different values of $\rm A/N$.}
\center
\begin{tabular}{c c c}
\hline\hline
$\sigma_{\rm N}$ & A/N & $Z_{\rm spu}$ \\
\hline
0.005 & 2 & 0.1499 \\
0.003 & 3.3 & 0.0987 \\
0.002 & 5 & 0.0612 \\
0.001 & 10 & 0.0305 \\
0.0007 & 14.3 & 0.0196 \\
0.0005 & 20 & 0.0174 \\
\hline
\end{tabular}
\label{ztab}
\end{table}

To quantify the effect of the spurious period variations, we calculate the parameter

\begin{equation}
Z = \frac{6\Delta P_{\rm w}}{P_{\rm w}}
\label{z}
\end{equation}

\noindent from the period estimates of the simulated data. This parameter measures the variability of $P$ within its weighted $\pm3\sigma$ limits. The definitions of the parameters are $w_i=\sigma_{P,i}^{-2}$, $P_{\rm w} = \sum{w_iP_i}\left[\sum{w_i}\right]^{-1}$ and $\Delta P_{\rm w} = \sqrt{\sum{w_i(P_i-P_{\rm w})^2}}/\sqrt{\sum{w_i}}$ \citep{jetsu1993decade}. The $Z$ values calculated from the independent datasets for all simulated models are given in Table \ref{ztab}. When using the same parameter $Z$ to characterise the differential rotation of a star, we may call these values spurious differential rotation $Z_{\rm spu}$.

What is striking in the results of Table \ref{ztab}, are the large values the spurious differential rotation $Z_{\rm spu}$, especially with lower $\rm A/N$. This severely affects the detectability of weak differential rotation. For example, real differential rotation causing variability of only 2\% in $P$ would be hard to distinguish from the spurious differential rotation even from data of good quality with $\rm A/N=20$. The values of $Z_{\rm spu}$ are roughly inversily proportional to A/N. Even very large 15\% variations of $P$ would be indistinguishable from the spurious differential rotation with $\rm A/N=2$.

By considering the period variations caused by physical and spurious differential rotation, i.e. $Z_{\rm phys}$ and $Z_{\rm spu}$, to be independent effects, we can estimate the contribution of the real physical differential rotation. In this case, the observed value of $Z^2$ is the sum of the squares of the two components, $Z^2=Z_{\rm phys}^2+Z_{\rm spu}^2$, yielding

\begin{equation}
Z_{\rm phys}^2 = Z^2 - Z_{\rm spu}^2.
\label{zphys}
\end{equation}

\noindent The value of $Z_{\rm spu}$ in this equation can be interpolated for any $\rm A/N$ based on the values in Table \ref{ztab}. In theory, Eq. \ref{zphys} gives an exact instantaneous value for $Z_{\rm phys}$ under the assumption that $Z_{\rm phys}$ and $Z_{\rm spu}$ are uncorrelated. Note, however, that both the amplitude $A$ of the light curve and the observational accuracy $\sigma_{\epsilon}$ are likely to change during a time series. Thus it is impossible to get a unique estimate for $Z_{\rm spu}$. Therefore, the relation of Eq. \ref{zphys} only gives a rough guideline for estimating the effect of the spurious period variations.

\section{Analysis of \object{HD 116956}}
\label{demo}

\begin{figure}
\resizebox{\hsize}{!}{\includegraphics[angle=90]{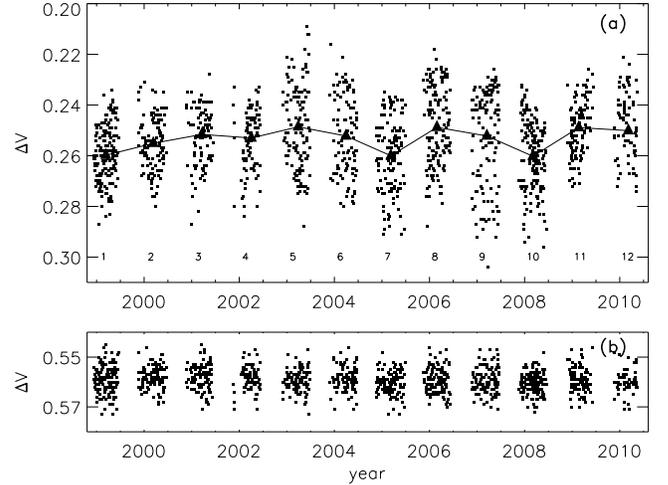}}
\caption{The differential photometry of (a) \object{HD 116956} 
and (b) the comparison star \object{HD 114446}. 
Each observing season is labeled with its segment number SEG=1,2,...,12. The filled triangles connected with continuous lines denote seasonal mean differential magnitudes.}
\label{vdata}
\end{figure}

In this section we present results of the CPS analysis of long-term photometry of the young solar analogue \object{HD 116956}. The analysis was done using the a priori period estimate $P_0=7.80$~d based on the preliminary results of \citet{gaidos2000spectroscopy}. The upper limit for the modelling order was set to be $K_{\rm lim}=2$.

To further justify the choice of $P_0$, we can estimate the lower limit for the stellar radius, $R\sin{i}=Pv\sin{i}/2\pi=0.86R_{\odot}$, by using $P=P_0$ and $v\sin{i}=5.6$ km s${}^{-1}$ as given in Sect. \ref{intro}. This limit is slightly smaller than the radius of the Sun, as expected for a star with a somewhat later spectral type. Another estimate for the stellar radius can be obtained using the Barnes-Evans relation \citep{lacy1977distances}

\begin{eqnarray*}
\log{(R/R_{\odot})} = 7.4724 - 0.2V_0 - 2F_V + \log{d},
\end{eqnarray*}

\noindent where $F_V=3.977-0.429(V-R)_0$ and $[d]=$ pc. Using $V_0=7.286$, $(V-R)_0=0.5$ and $d=21.9$ pc from the Hipparcos \citep{perryman1997hipparcos} and USNO-B \citep{monet2003usno} catalogues and neglecting the interstellar reddening because of the small distance to the star, we get $R=(0.66\pm0.06)R_{\odot}$. This value is slightly smaller than the lower limit calculated using the rotation period and velocity. Nevertheless, the two results for the radius are of the same order and $P_0$ is a reasonable estimate for the rotation period.

Our photometry\footnote{The V band photometry used in this paper is published electronically at the CDS.} of \object{HD 116956} was obtained over 12 consecutive observing seasons between $\rm HJD=2451172$ (24 December 1998) and $\rm HJD=2455342$ (25 May 2010) with the T3 0.4 m automatic photoelectric telescope (APT) at Fairborn Observatory in Arizona. The APT performs differential photometry in the Johnson $V$ and $B$ passbands. A brief description of the operation of the APT and the reduction of the data can be found in \citet{fekel2005chromospherically} and references therein. A total of 1408 differential $V$ band observations were acquired with \object{HD 114446} as the comparison star. This data is shown in Fig. \ref{vdata}, which also shows the season numbers that coincide with the segment division of the CPS analysis. The seasonal means are denoted with filled triangles which have been connected with continuous lines.

The external precision of the observations taken with the T3 telescope is $0.004-0.005$ mag, as shown by simultaneous measurements of the comparison star \object{HD 114446} against a check star \object{HD 119992} (see Fig. \ref{vdata} (b)). The difference between the comparison and check star showed no seasonal changes during the whole observing period, i.e. there are no prominent systematic errors in the data. This is also supported by the $\chi^2$-test for the $n=1273$ comparison star minus check star differential magnitudes. Using $\epsilon_{\rm i}=y_{\rm i}-m_{\rm y}$ and $\sigma_i=0.005$ gives $\chi^2=1166.6$ (Eq. \ref{chi2}). This corresponds to the critical level $Q=0.9829$, where $Q$ states the probability of $\chi^2$ reaching the observed value under the null hypothesis that these differences remain constant. There is no need to reject this constant brightness hypothesis. In contrast, the $n=1408$ \object{HD 116956} minus comparison star differential magnitudes give $\chi^2=12002.9$ which corresponds to $Q \ll 10^{-20}$, i.e. these observations certainly exhibit variability. For a more detailed description of the data acquisition see \citet{henry1999techniques}.

\subsection{Graphical presentation of the results}
\label{graph}

\begin{figure*}
\centering
\includegraphics[angle=90,width=17cm]{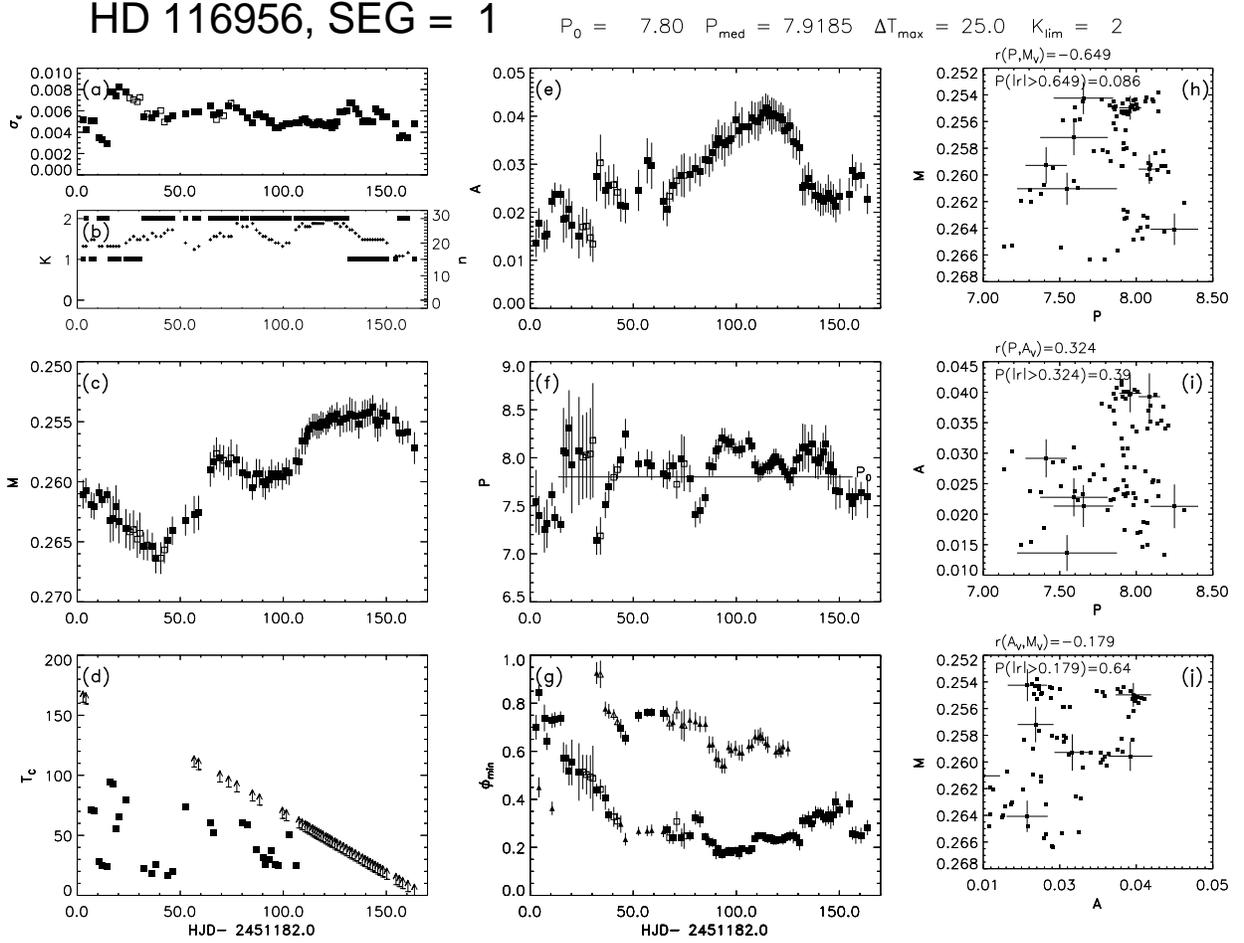}
\caption{The CPS analysis of $\rm SEG=1$ of the photometry of \object{HD 116956}. Descriptions of the subplots are given in Sect. \ref{graph}.}
\label{seg1}
\end{figure*}

The CPS divided the data into twelve segments, each constituting of a complete observing season. The observing seasons typically lasted from the beginning of December to the beginning of July. The first season of our data has also been analysed by \citet{gaidos2000spectroscopy}. The analysis of this first segment ($\rm SEG=1$) is presented graphically in Fig. \ref{seg1}, where the units are V magnitudes for $\sigma_{\epsilon}$, $M$ and $A$ and days for $T_{\rm C}$ and $P$. All the rest of the parameters are dimensionless. This figure contains the subplots, where the parameters are plotted as functions of $\tau$:

\begin{itemize}
\item[{\bf (a)}] Standard deviation of residuals $\sigma_{\epsilon}(\tau)$
\item[{\bf (b)}] Modelling order $K(\tau)$ (squares, units on the left y-axis) 
and number of observations per dataset $n$ (crosses, units on the right y-axis)
\item[{\bf (c)}] Mean differential V-magnitude $M(\tau)$ (HD 116956 minus HD 114446
\item[{\bf (d)}] Time scale of change $T_{\rm C}(\tau)$
\item[{\bf (e)}] Amplitude $A(\tau)$
\item[{\bf (f)}] Period $P(\tau)$
\item[{\bf (g)}] Primary (squares) and secondary (triangles) minimum phases $\phi_{\rm min,1}(\tau)$ and $\phi_{\rm min,2}(\tau)$
\item[{\bf (h)}] $M(\tau)$ versus $P(\tau)$
\item[{\bf (i)}] $A(\tau)$ versus $P(\tau)$
\item[{\bf (j)}] $M(\tau)$ versus $A(\tau)$.
\end{itemize}

\noindent In the subplots (a), (c) and (e)--(g), the reliable parameter estimates are indicated by filled symbols and unreliable ones by open symbols. The reliability of the parameter estimates is determined as described in Sect. \ref{modelling}. In the case of $P$, also the level of the a priori period estimate $P_0$ is shown (horizontal line). This fits reasonably well to the detected $P$ values since the same data of $\rm SEG=1$ was used by \citet{gaidos2000spectroscopy} to determine $P_0$. The minimum phases are calculated using the median period $P_{\rm med}$ of the segment. In subplot (d), the upward pointing arrows indicate that the computation of the time scale of change has reached the end of the segment and they correspond to the values $T_{\rm C}=-2$ (see App. \ref{format}). In the correlation plots (h)--(j), the error bars have been drawn only for the parameter estimates of independent datasets. The linear Pearson correlation coefficients $r_0$ for the independent datasets, as well as the probabilities $P(|r|>r_0)$, are given. Finally, the values of the a priori period estimate $P_0$, the median period $P_{\rm med}$, the limiting modelling order $K_{\rm lim}$ and the maximum length the a dataset $\Delta T_{\rm max}$ are given above the plot.

\begin{figure*}
\centering
\includegraphics[angle=90,width=17cm]{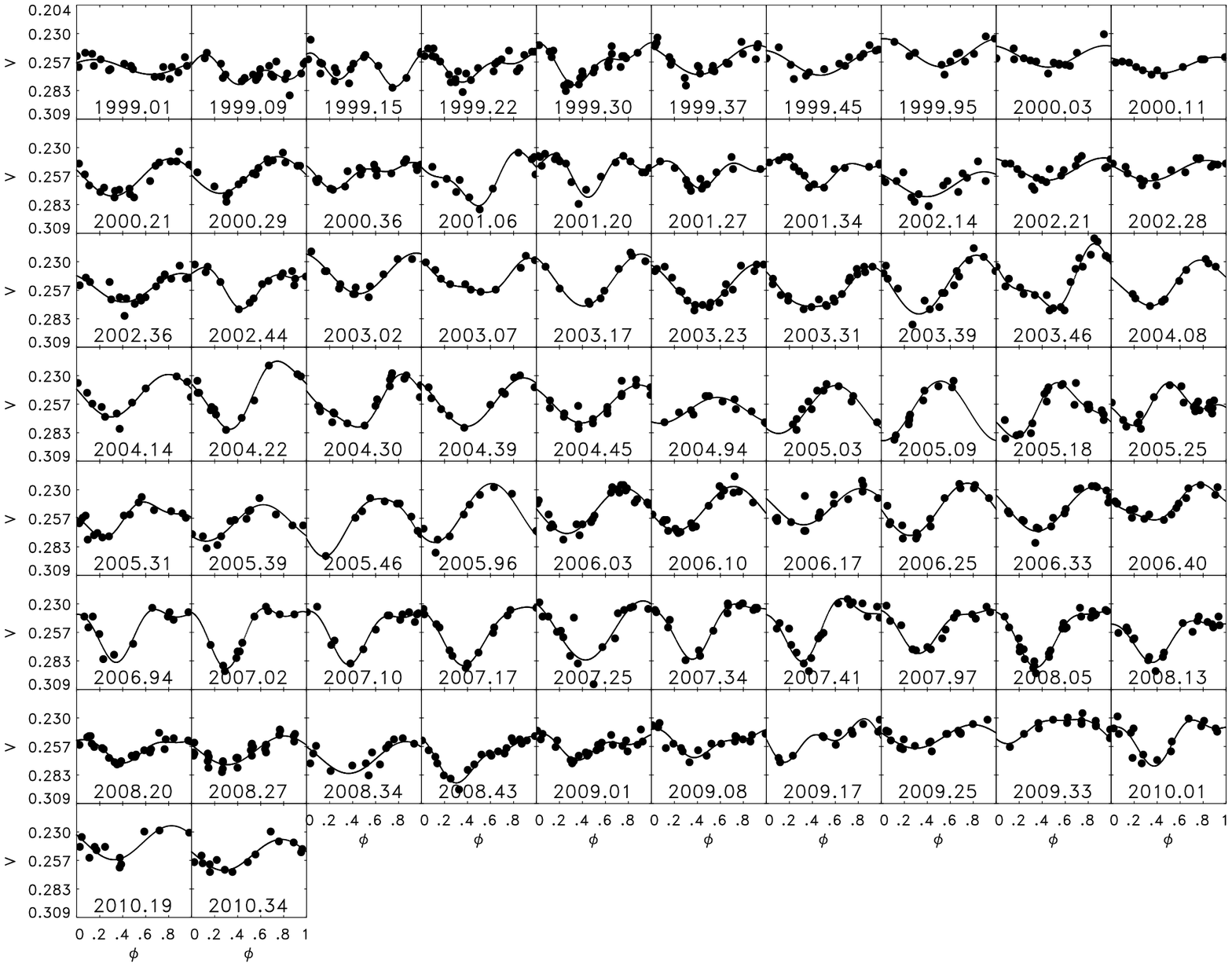}
\vspace{-10mm}
\caption{Light curves of independent datasets: Each dataset was first modelled with the phases $\phi_1={\rm FRAC}[(t-t_{\rm min,1}(\tau))/P(\tau)]$. The phases $\phi_{\rm al,1}$ of the primary minimum epochs $t_{\rm min,1}(\tau)$ were then computed with the constant period the ephemeris ${\rm HJD}_{\rm al}=2451176.94+7.8416E$ (see Fig. \ref{aktlong}). Finally, the data and the light curves were plotted as a function of the phase $\phi=\phi_1+\phi_{\rm al,1}$.}
\label{curvefit}
\end{figure*}

The light curve fits for the independent datasets are presented in Fig. \ref{curvefit}. Because the CPS models each of the datasets with a different period $P(\tau)$, it is in principle impossible to represent these light curves with an ephemeris based on one constant period. Each dataset was therefore first modelled with the phase $\phi_1={\rm FRAC}[(t-t_{\rm min,1}(\tau))/P(\tau)]$, i.e. a different period value was used for each dataset. Then the phases $\phi_{\rm al,1}$ of the primary minimum epochs $t_{\rm min,1}$ were computed with the constant period ephemeris ${\rm HJD}_{\rm al}=2451176.94+7.8416E$ (see Sect. \ref{al}). Finally, the data and the light curve models were displayed as a function of the phase $\phi=\phi_1+\phi_{\rm al,1}.$ We chose this particular definition for the phase, because it allows an easier comparison between Figs. \ref{curvefit} and \ref{aktlong}.

\subsection{Long term variations}

\begin{figure}
\resizebox{\hsize}{!}{\includegraphics[angle=90]{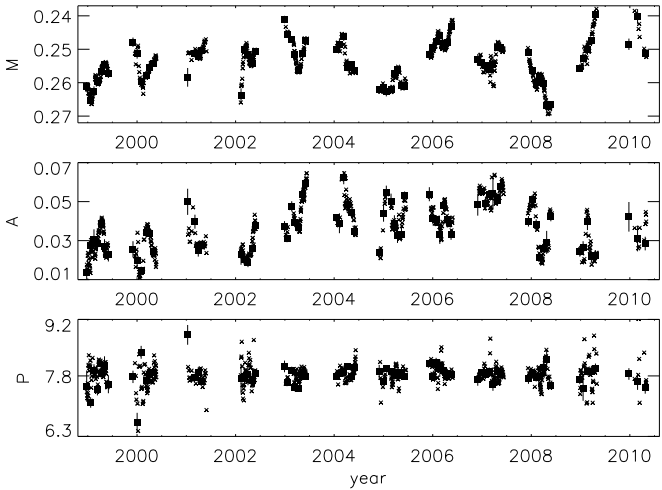}}
\caption{Evolution of $M$, $A$ and $P$ during the complete time series. Parameter estimates from independent datasets are denoted as squares with error bars and from all other datasets as small crosses. $M$ and $A$ are given in V magnitudes and $P$ in days.}
\label{sykli}
\end{figure}

Our Fig. \ref{sykli} displays the long term variation of the light curve parameters $M$, $A$ and $P$. With the exception of the rotation period $P$, these seem to display some regular behaviour. This may be a sign of activity cycles in the star. The variations of $P$ seem to be more or less random.

The long term variations are most striking in the mean brightness $M$. During 2004--2005 it underwent a dip of $\sim0.02$ mag. A similar dip occurred later in 2008. Also at the beginning of the observations in 1999, the values of $M$ were dimmer than the average suggesting another similar dip. The same variability can be seen in the raw $V$ data (Fig. \ref{vdata}). In addition to the minima before 1999 and in 2005 and 2008, there seems to have been one during 2001--2002. This one cannot be seen in the $M$ variations. It is possible that signs of this particular dip are lost in the gap between the observing seasons of 2001 and 2002. Put together, these timings might suggest an activity cycle of roughly 3 years in length.

The variations of the mean brightness of the star have the most straight forward interpretation as an activity proxy. As the spottedness of the star increases, its mean effective temperature decreases and a lower brightness is observed. Another proxy is provided by the light curve amplitude $A$. It measures more or less the nonaxisymmetry of the spot distribution. There may or may not be a correlation between this and the total spottedness and likewise between the variations of $A$ and $M$. As it happens, the connection between the long term variations of $M$ and $A$ is not immediately clear. There seem to be variations in $A$ with about the same time scale as in $M$. However, these variations are more erratic than those of $M$.

To investigate somewhat further, whether the star has any activity cycles, we performed the CPS analysis for the independent $M$ and $A$ estimates, using $\sigma_{M}^{-2}$ and $\sigma_{A}^{-2}$ respectively as weights. Our a priori cycle period estimate was $P_0=3.0$ yrs. The resulting cycle periods were $P_{\rm M}=3.26\pm0.04$ yrs and $P_{\rm A}=3.09\pm0.13$ yrs. However, these models had $\chi^2_{\rm M}=1206.2\gg n_{\rm ind}=72$ and $\chi^2_{\rm A}=847.1\gg n_{\rm ind}=72$, where $n_{\rm ind}$ is the total number of the independent datasets. Any reasonable model should satisfy $\chi^2 \approx n_{\rm ind}$, i.e. these two cycles are simply not statistically significant. Thus, the data does not seem to support the interpretation of the seasonal $M$ and $A$ variations as signs of activity cycles.

There may be short time scale correlations between $M$, $A$ and $P$. For $\rm SEG=1$, the linear Pearson correlation coefficients are $r_0(P,M)=-0.649$, $r_0(P,A)=0.324$ and $r_0(A,M)=-0.179$. Because of the small number of independent datasets ($n_{\rm ind}=7$) none of these are significant. The same applies for all other segments as well. Other than simply linear correlations are also possible, though even more difficult to detect. The presence of such correlations is suggested by paths of parameter estimate pairs in Figs. \ref{seg1} (h)-(j).

\subsection{Active longitudes}
\label{al}

\begin{figure}
\resizebox{\hsize}{!}{\includegraphics[angle=90]{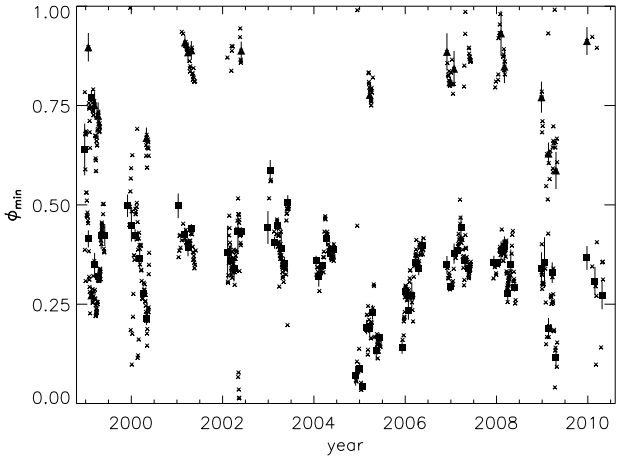}}
\caption{Minimum phases $\phi_{\rm min,1}$ and $\phi_{\rm min,2}$ with the constant period ephemeris ${\rm HJD}_{\rm al}=2451176.94+P_{\rm al}E$, where $P_{\rm al}=7.4816$ days. Estimates from independent datasets are denoted as squares (primary minima) and triangles (secondary minima) with error bars. Estimates from the rest of the datasets are denoted with small crosses.}
\label{aktlong}
\end{figure}

The long term variation of the primary and secondary light curve minimum phases is displayed in Fig. \ref{aktlong} with the same notation as in Fig. \ref{sykli}. The minimum phases $\phi_{\rm min,1}$ and $\phi_{\rm min,2}$ are calculated from the minimum epochs $t_{\rm min,1}$ and $t_{\rm min,2}$ using the period $P_{\rm al}=7.8416\pm0.0011$ d. This period was detected when the Kuiper test was applied to the $n=90$ reliable primary and secondary minimum epoch estimates of all independent datasets.

The formulation of this nonweighted K-method was the same as in \citet{jetsu1996searching}. It tests the null hypothesis ($\rm H_0$) that the phases of the epochs of the primary and secondary minima are evenly distributed between 0 and 1. Under $\rm H_0$, one can determine the probability that the K-method test statistic reaches any particular value within the chosen tested period interval. This probability is called the critical level $Q_{\rm K}$ and if it is very small, the null hypothesis must be rejected. The best detected period has the smallest $Q_{\rm K}$.

The period interval that was tested with the K-method was between $P_{\rm min}=0.85P_0=6.63$ d and $P_{\rm max}=1.15P_0=8.97$ d. The critical level of the best 7.8416 d periodicity was extremely significant, $Q_{\rm K}=8.7\cdot10^{-11}$. It exceeds all critical level estimates given in Table 2 by \citet{jetsu1996active}, where active longitudes were detected in \object{$\lambda$ And}, \object{$\sigma$ Gem}, \object{II Peg} and \object{V711 Tau}. Such an extreme value of $Q_{\rm K}$ confirms the presence of two long-lived active longitudes in \object{HD 116956}.

A striking feature of the minimum phases is that they are confined to two active longitudes with a phase separation of $\Delta\phi\approx0.5$. Comparing Figs. \ref{aktlong} and \ref{seg1} (g) shows that this phenomenon is present in both long and short term evolution of the light curve. Considering the results of the analysis in Sect. \ref{testmin}, the existence of two confined areas of light curve minimum phases could be attributed to a selection effect. However, both the long term stability and the fact that there is very little short term random changes in the minimum phases show that the observed active longitudes are indeed likely to be real. If they were created by a selection effect, they would not remain stable on time scales longer than the length of single datasets $\Delta T_{\rm max}$.

The existence of a stable period with which the active longitudes rotate indicates that there is some coherent magnetic structure inside the star rotating with the period $P_{\rm al}$. Such a structure could be generated by a nonaxisymmetric dynamo mode \citep[p.271]{krause1980meanfield}.

In addition to the long term stability, the active longitudes seem to have experienced periods of migration. This is evident especially from the seasonal primary minimum phases. Between the observing seasons of 2004 and 2005, there was a phase jump of $\Delta\phi\approx0.3$. During the next years, the primary minimum migrated to the same phase where it had resided before the phase jump. Quite remarkably, this transient occurred at the same time as the minimum of $M$ at 2005. It is not clear whether these two phenomena are related to each other. Another short phase jump seems to have occurred at 2009 and possibly one also at 2000. In both cases, the primary and secondary minima seemed to experience the same phase shift with their mutual phase separation staying more or less constant.

Another remarkable property in the distribution of the minimum phases is that for most of the time the primary minima are confined to one active longitude and the secondary minima to the opposite one. This is evident in Fig. \ref{aktlong} where the primary minima present in the independent datasets are denoted as squares and the secondary minima as triangles. Only at the beginning of the time series are there exceptions to this. Two of the independent datasets had their primary minima near the phases where the following independent datasets had their secondary minima and one had its secondary minimum near the phases where the other independent datasets had their primary minima. The course of events can more closely be examined in Fig. \ref{seg1} (g). At the beginning of the first segment, the primary light curve minima were located near $\phi=0.7$. As the segment continued, there was a migration of observed primary minimum phases to $\phi=0.2$, where a new active longitude formed. Roughly 70 d after the start of the segment, there was a shift in the order of the minima so that the secondary minimum grew deeper and became the new primary minimum. Towards the end of the segment, the old primary minimum gradually faded away. This change from a primary minimum to a secondary and vice versa if a direct proof of the occurrence of a flip-flop event in \object{HD 116956} \citep{jetsu1993spot}.

\begin{figure}
\resizebox{\hsize}{!}{\includegraphics[angle=90]{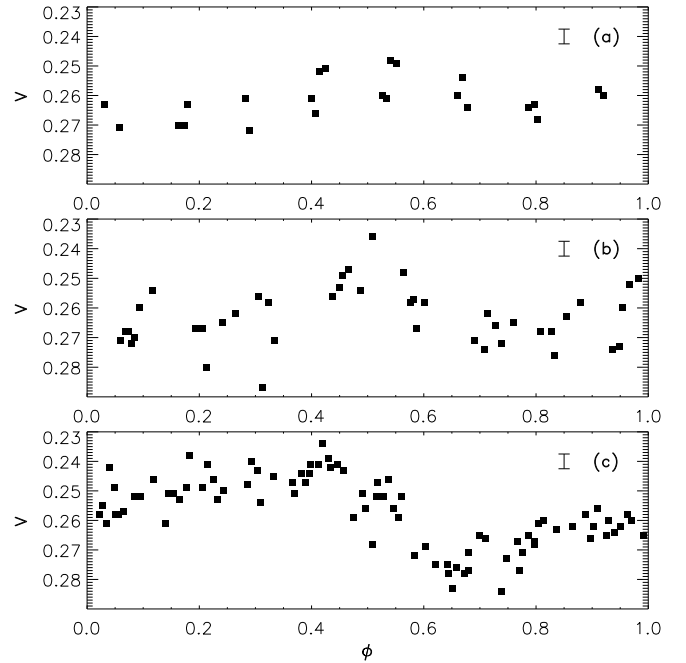}}
\caption{The light curve of HD 116956 during $\rm SEG=1$ folded with median period $P_{\rm med}=7.8896$ d and divided into three parts. Panel (a) includes observations from the first 29 nights of the segment, panel (b) observations from the subsequent 56 nights and panel (c) observations from the last 104 nights. The scale of the 1$\sigma$ accuracy of the data (0.005 mag) is indicated with a separate error bar at the right upper corner of the panels.}
\label{seg1lc}
\end{figure}

Another closer look at the flip-flop in segment 1, that is observations between $\rm HJD=2451172$ and $\rm HJD=2451361$, is provided by Fig. \ref{seg1lc}. It displays the observations folded with the median period $P_{\rm med}=7.8896$ d of the segment. Because of period variations during the segment, the resulting light curves are blurred. Nevertheless, the general variations of the light curve shape remain clearly visible. The observations in Fig. \ref{seg1lc} are divided into three parts, panel (a) includes observations from the first 29 nights of the segment, panel (b) observations from the subsequent 56 nights and panel (c) observations from the last 104 nights. During the first and second parts of the segment, the light curve displayed two minima near $\phi=0.2$ and $\phi=0.8$ and its total amplitude remained relatively low. A distinguishing feature between the two parts is the relation between the depths of the two minima. During the first part, the minimum near $\phi=0.2$ is distinctly the deeper of the two. During the second part, no clear differences between the depths are visible. Quite remarkably, there was an abrupt change in the light curve shape between the second and third parts. The minimum near $\phi=0.8$ was slightly shifted and the minimum near $\phi=0.2$ disappeared. Also the light curve amplitude increased at the same time. As the observations continued unbroken during the changes, we must conclude that the changes were relatively fast. At the same time with the changes in the light curve shape there was a jump in $M$ (Fig. \ref{seg1} (c)), which may have been connected to this flip-flop event.

The phase migrations at 2005 and 2009 may not be interpreted as flip-flops in the sense of \citet{jetsu1993spot}, as there were no switches of activity between the two active longitudes. Rather, both the primary and secondary minima remained at their own active longitudes, as well as preserved their phase separation.

\subsection{Differential rotation}
\label{diffrot}

If the stellar photosphere rotates differentially, we can expect to observe different rotation periods for spots at different latitudes. At any particular time, only one value of $P$ can be observed. But as the spot distribution evolves, so should the observed rotation period. The observed period values correspond to the total contribution of all the spots on the star weighted by their contrast to the unspotted surface and their visibility.

As already described in Sect. \ref{teststab}, we can measure the strength of the surface differential rotation using the $\pm3\sigma$ limits of $P$ variations (Eq. \ref{z}). Using only period estimates from the independent datasets, we get the weighted mean period $P_{\rm w}\pm\Delta P_{\rm w}=7.8288\pm0.1423$ d and the strength of the variations $Z=0.11\equiv11\%$.

Combining the observational accuracy of 0.005 mag and a typical light curve amplitude $A=0.05$ mag gives a signal to amplitude ratio $\rm A/N=5$. Using the results of the analysis in Sect. \ref{teststab}, this $\rm A/N$ value would introduce spurious period variations of $Z_{\rm spu}=0.06\equiv6\%$. We can investigate the contribution of this effect by calculating a rough estimate for the physical component of the period variations using Eq. \ref{zphys}. This gives $Z_{\rm phys}\approx0.09\equiv9\%$, which is still larger than $Z_{\rm spu}$, although being of the same order of magnitude. Thus there remains a possibility that even these period variations are not physical and no differential rotation is observed. The possibility of a constant period in the observational data can, however, be rejected by computing the $\chi^2$ value (Eq. \ref{chi2}) of the $n_{\rm ind}=72$ independent period estimates against the hypothesis $P=P_{\rm w}\equiv \rm constant$. The residuals $\epsilon_{\rm i}=P_{\rm i}-P_{\rm W}$ and the errors $\sigma_{\rm P_i}$ give $\chi^2=282$, which corresponds to a critical level $Q \ll 10^{-20}$, i.e. the hypothesis of a constant period must be rejected.

Assuming that the period variations are indeed caused by differential rotation, we can try to interpret the $Z$ value in somewhat more depth. If we make the typical assumption of solar like differential rotation profile $P(b)=P_{\rm eq}/(1-k\sin^2{b})$, where $b$ is latitude and $P_{\rm eq}$ the equatorial rotation period, we can relate the calculated $Z$ value to the differential rotation coefficient $k$ with the scaling law $|k|\approx Z/h$, where $h=\sin^2{b_{\rm max}}-\sin^2{b_{\rm min}}$ and $b_{\rm min}$ and $b_{\rm max}$ are the minimum and maximum latitudes of spot formation \citep{jetsu2000time}. In order to do this we must, however, know the latitudinal extent $[b_{\rm min},b_{\rm max}]$ of the spot activity. If the spots are restricted to a narrow latitude interval, we get observational data only from the rotation periods within that region. The actual value of $k$ is thus larger than what $Z$ suggests. If the spot activity spans all the way from the equator to the poles, the scaling coefficient $h$ approaches unity and $Z\approx |k|$. However, spots near the poles contribute less, or not at all, to the rotational modulation of brightness. This means that the suitable values of $h$ are always somewhat less than unity.

In the case of \object{HD 116956} we have no knowledge of the latitude extent of the spot activity. We can, however, apply different scaling factors. In the case of total range of spot activity from $b_{\rm min}=0^{\circ}$ to $b_{\rm max}=90^{\circ}$, we have $h=1$ and $|k|=0.11$. This is roughly half of the solar value of $k=0.20$. But for a solar like distribution of spot activity from $b_{\rm min}=0^{\circ}$ to $b_{\rm max}=30^{\circ}$, we get $h\approx0.25$ and $|k|=0.44$, over twice the solar value. Such a large value renders the assumption of solar like spot distribution highly dubious. It would be more likely that the spot activity on HD 116956 is distributed over a larger latitudinal range than on the Sun and the value of $k$ be closer to $k=0.11$. This would correspond to differential rotation rate $\Delta\Omega=k\Omega=2\pi k/P=0.088$ rad d${}^{-1}$ between the equator and the poles.

\citet{henry1995starspot} investigated the relation between the observed values of $k$ and the stellar rotation periods $P_{\rm rot}$ and arrived at the relation

\begin{equation}
\log{k} = -2.12 + 0.76\log{P_{\rm rot}} - 0.57F,
\label{kp}
\end{equation}

\noindent where $P_{\rm rot}$ is in days. The Roche lobe filling factor $F=R_{\rm star}/R_{\rm Roche}$ reduces to $F=0$ for single stars. Using the weighted mean period $P_{\rm w}=7.8288$ d we get a prediction $k=0.036$ for \object{HD 116956}. This is a low value, about one third of the observed lower limit $k\geq0.11$. Thus it seems that, both the lower and upper limits $k=0.11$ and $k=0.44$ reside in or near the scattered region of differential rotation estimates in Fig. 28 of \citet{henry1995starspot}, and therefore \object{HD 116956} would appear to undergo much stronger differential rotation than expected.

Another comparison was done using the results presented by \citet{collier2007differential}. Their relation between differential rotation rate ($\Delta\Omega$) and the effective temperature of the star ($T_{\rm eff}$)
was

\begin{equation}
\Delta\Omega = 0.053\left(\frac{T_{\rm eff}}{5130}\right)^{8.6},
\end{equation}

\noindent where $T_{\rm eff}$ is in kelvins and $\Delta\Omega$ in radians per day. This predicts $\Delta\Omega=0.057$ rad d${}^{-1}$ for \object{HD 116956}. This is again smaller than the observed value, although closer to it than what was predicted earlier by Eq. \ref{kp}.

\subsection{Time scale of change}

The parameter $T_{\rm C}$ is an estimate of the typical time scale in which the spot distribution changes. It is an important parameter when investigating the short term dynamics of the spot activity. Unfortunately, interpreting the values of $T_{\rm C}$ is not simple. As can be seen in Fig. \ref{seg1} (d), the $T_{\rm C}$ values evolve quite rapidly. Moreover, there are a large number of datasets where the model has been applicable right to the end of the segment (denoted with arrows), i.e. no $T_{\rm C}$ estimate has been obtained.

The seemingly random evolution of $T_{\rm C}$ stems from many causes. First of all, the complexity of the model affects the time that this same model stays applicable to future observations. Simple models with low $K$ tend to have longer $T_{\rm C}$ than more complex ones. This is apparent from the very first few datasets of $\rm SEG=1$, as they have $K=1$ (Fig. \ref{seg1} (b)) and $T_{\rm C}>160$ d (Fig. \ref{seg1} (d)) reaching the end of the segment.

An incomplete phase coverage of the light curve can also cause a too low value of $T_{\rm C}$. If there is a phase gap in the light curve with no observations, there may be a situation where the step to the next dataset introduces new observations to the phase gap thus completing the phase coverage. In this case, the contribution of new observations can significantly alter the light curve so that the model of the previous dataset is no longer applicable and $T_{\rm C}$ gets a very low value.

Even if there are no significant phase gaps in the observations, an increasing amount of observational data can affect the value of $T_{\rm C}$. This is because with more data the light curve can be determined more accurately and there is a higher probability of choosing a more complex model. This leads to smaller values for $T_{\rm C}$. Through a concrete example, it can be seen how a larger number of observations and a better observational accuracy decrease the values of $T_{\rm C}$. These effects can be nicely seen in the observations of \object{$\epsilon$ Eri} by \citet{croll2006differential}. They analysed continuous photometry of the star taken with the MOST satellite during three rotation periods. Their light curve is both well sampled and accurate and it turns out to have a different shape during each of the three rotations, as can be seen in their Fig. 2.

Because of the complex behaviour of $T_{\rm C}$, it is best to look at the long-term mean time scale of change rather than studying the short term changes. This long-term mean for \object{HD 116956} is $\overline{T}_{\rm C}=44.1$ d. This value is nearly two times as large as the length of the datasets $\Delta T_{\rm max}=25$ d. Thus the light curve usually remains unchanged during one typical dataset and the chosen value of $\Delta T_{\rm max}$ is reasonable.

The timescale of light curve change can be compared to the convective turnover time $\tau_c$, as mixing of the convection zone may alter the spot distribution which is then observed as a change in the light curve. \citet{kim1996theoretical} performed theoretical modelling to find the values of $\tau_c$ in late type stars. \citet{ossendrijver1997cycle} published an interpolation formula for their results by fitting a cubic polynomial to the theoretically calculated values,

\begin{equation}
\tau_c = -68.3 + 224.8x - 177.2x^2 + 57.0x^3,
\end{equation}

\noindent where $x=B-V$. This formula gives $\tau_c$ in days. By using the Hipparcos value of $B-V=0.823$ for \object{HD 116956} \citep{perryman1997hipparcos}, we get $\tau_c=28.5$ d. Even if the two values are of the same order of magnitude, it is not clear whether there is a meaningful link between $\tau_c$ and $\overline{T}_{\rm C}$.

\section{Conclusions}

We have formulated a new Continuous Period Search (CPS) method by improving the earlier TSPA method described in Paper I. Our goal was to develop better tools for the analysis of photometry of active stars. The three new features are:

\begin{itemize}
\item[(1)] A sliding window for selecting overlapping datasets.
\item[(2)] A criterion for selecting the correct order for the model.
\item[(3)] A time scale for the change $T_{\rm C}$ of the model.
\end{itemize}

\noindent Using the sliding window significantly improves the time resolution of the analysis. The CPS determines the light curve parameters with a time resolution much shorter than the length of individual datasets. In contrast, the TSPA could only achieve a time resolution equal to the dataset length. A similar sliding window was applied for a first order harmonic model in \citet[their Eq. 3]{berdyugina2002magnetic}. Our model order selection criterion allows the selection of a model with suitable complexity for describing the data in each individual dataset. This even allows the alternative of a constant light curve with no periodicity. Since the CPS can use Fourier series of any arbitrary order as a model, it can in principle be applied to any type of periodic data. Finally, the computed $T_{\rm C}$ parameter measures the time scale of change of the light curve.

In order to characterise the performance of the CPS method, we tested it with simulated data. We found imperfect performance in three critical situations.

Firstly, the model order selection criterion does not give the correct order $K$ in all of the analysed datasets. This is caused by the limited number of observations. Typically, when analysing photometry of active stars, each dataset contains $10 \leq n \leq 30$ observations. The success rate in finding the correct modelling order depends on $n$.  In the case of analysing periodic data, the success rate is between 80\% and 90\% for datasets with $n\leq 20$ and larger than 90\% for datasets with $n>20$. In the case of constant noisy data with no periodicity, this success rate is much smaller, and spurious periodicity is still detected in 40\% and 20\% of the datasets having $n \leq 20$ and $n>20$, respectively. This reduces our ability to distinguish between time intervals of periodic variation and constant brightness in the light curve. However, this type of spurious periodicity can often be identified from the low observed amplitude to noise ratio of the models.

Secondly, it is not always possible to uniquely detect two separate minima from the light curves of spotted stars. This is a general problem and common to all time series analysis methods. Below some critical phase separation $\Delta\phi_{\rm crit}$, the two minima merge into one common minimum. We calculated a theoretical limit of $\Delta\phi_{\rm crit}=0.25$ and then obtained $\Delta\phi_{\rm crit}\approx0.33$ from simulated data analysed with the CPS. The existence of this type of a phase limit is a concern when analysing photometry of active stars, since it means that two individual spots must have a considerable longitudinal separation to be detected separately. If one is not careful, this may lead to detections of spurious active longitudes.

Thirdly, there is an intrinsic instability in the period estimation. This is a concern because period variations are often used to measure the differential rotation of active stars. For good quality data the effect is limited, but grows substantially for more noisy data. This period instability is caused by random effects due to noise, which is a general problem for all time series analysis methods, especially when the analysed datasets are short.

We applied the CPS to 12 years of $V$ band photometry of the young solar analogue \object{HD 116956}. The analysis revealed variations in the mean magnitude $M$ and the light curve amplitude $A$ with an apparent period around 3 years. However, we found no conclusive evidence supporting the interpretation that these variations are signs of an activity cycle, as the CPS analysis for the $M$ and $A$ estimates gave very high $\chi^2$ values for a periodic fit.

The star also displays two active longitudes that have remained stable over the whole 12 year observing period. The standard Kuiper test of the primary and secondary minimum epochs gave an extreme critical level $Q_{\rm K}=8.7{\rm x} 10^{-11}$ for  the rotation period $P_{\rm al}=7.8416\pm0.0011$~d of these active longitudes. There have been only few transient excursions to other longitudes and the separation of the primary and secondary minima has stayed nearly unchanged. During the first observing season at 1999, the star underwent a flip-flop. In this event, the major activity switched from one active longitude to the other, while the old primary minimum disappeared completely. This flip-flop event happened very fast.

We estimated the differential rotation of the star, assuming that it is the cause of the observed photometric period variations. These $\pm3\sigma$ variations of the period gave the value $Z=11\%$, which is much larger than what could be caused only by spurious period variations. Assuming that the spot distribution covered the whole latitude range from equator to pole and that the solar law of differential rotation were valid also for \object{HD 116956}, these variations would correspond to a differential rotation coefficient $|k|=0.11$, or equivalently a differential rotation rate $\Delta\Omega=0.057$ rad d${}^{-1}$. The observed solar value is $k=0.20$. If the latitudinal extent of the spot activity in \object{HD 116956} were the same as in the Sun, surface differential rotation would be much stronger, i.e. $|k|=0.44$.

The mean time scale of change of the light curve, i.e. the spot distribution of the star, was $\overline{T}_{\rm C}=44.1$ d. This exceeds the length of the datasets, $\Delta T_{\rm max}=25.0$ d. Hence the spot distribution remained unchanged during a typical dataset. There may be a link between the above $\overline{T}_{\rm C}$ value and the convective turnover time $\tau_c=28.5$ d, having the same order of magnitude. This could be expected, if convective mixing causes the observed changes in the distribution of starspots.

\begin{acknowledgements}
The work by P. Kajatkari was supported by the Vilho, Yrj\"{o} and Kalle V\"{a}is\"{a}l\"{a} Foundation.

The automated astronomy program at Tennessee State University has been supported by NASA, NSF, TSU and the State of Tennessee through the Centers of Excellence program.
\end{acknowledgements}

\begin{appendix}
\section{Format of the results}
\label{format}

Through the CPS method we obtain a lot of information useful for studying stellar activity. These results are published electronically and it is therefore necessary to describe the notation format in detail. The results of each individual dataset have been compressed to nine lines of an ASCII file. The parameter estimates are written in the format given in Table \ref{coding}. This table also specifies the units of the parameters.

The first row in Table \ref{coding} gives the segment number (SEG), the epoch of the first observation in the segment ($t_0$) and the dataset number (SET). The second row gives the epochs of the first ($t_1$) and the last ($t_n$) observation in the subset, the mean epoch of the dataset ($\tau$) and specifies whether the dataset is considered an independent dataset or not (IND). For independent datasets the value is IND=1, otherwise IND=0. The third row gives the number of observations ($n$), their mean ($m_y$) and their standard deviation ($s_y$). The fourth row gives the order of the model ($K$), the standard deviation of the residuals ($\sigma_{\epsilon}$) and the time scale of change ($T_{\rm C}$). As already mentioned in Sect. \ref{correlation}, the value $T_{\rm C}=-2$ indicates that the model describes well all the remaining datasets in the segment. The rest of the rows give values for the light curve parameters ($M$, $P$, $A$, $t_{\rm min,1}$, $t_{\rm min,2}$) and their errors ($\sigma_M$, $\sigma_P$, $\sigma_A$, $\sigma_{t_{\rm min,1}}$, $\sigma_{t_{\rm min,2}}$). It is also specified whether these parameter estimates are considered reliable or not. Reliable estimates have R=0, while unreliable ones have R=1. All the heliocentric Julian dates are given in the truncated form $\rm HJD-2\ 400\ 000$. If for some reason no value has been obtained for some parameter, the "dummy" value $-1$ is given.

A typical example of an entry into the output file is:

\vbox{
\begin{verbatim}
           1  51171.9835          17
  51200.9434  51225.8738  51214.2168       1
          22      0.2658      0.0086
           2      0.0055     22.2344
      0.2654      0.0012           0
      7.1382      0.1267           0
      0.0274      0.0045           0
  51207.7204      0.1689           0
  51203.6602      0.2811           1
\end{verbatim}
}

\noindent This is the 17th dataset in the first segment (SEG=1, SET=17) in the analysis\footnote{The full results of the analysis of HD 116956 are published electronically in this format at the CDS.} of the star HD 116956 presented in Sect. \ref{demo}. It is a dataset of $n=22$ observations obtained during a period of about $\Delta T=t_n-t_1=25$ d centered around $\rm HJD=2451214.22$. The dataset has been labelled as an independent one (IND=1). The modelling has been done with $K=2$ and the values for all model parameters $M$, $P$, $A$, $t_{\rm min,1}$ and $t_{\rm min,2}$ have been obtained. Only the $t_{\rm min,2}$ estimate is found to be unreliable ($R=1$).

Another example is the 11th dataset in the second segment (SEG=2, SET=11):

\vbox{
\begin{verbatim}
           2  51515.0410          11
  51560.9421  51585.8988  51574.4281       0
          13      0.2561      0.0084
           1      0.0063     -1.0000
      0.2568      0.0019           1
      8.3355      0.5855           1
      0.0154      0.0046           1
  51567.3488      0.7673           1
     -1.0000     -1.0000           1
\end{verbatim}
}

\noindent This dataset is not independent (IND=0). It contains only $n=13$ observations and has been modelled with a simpler $K=1$ model. For this model, there does not exist a secondary minimum and correspondingly $t_{\rm min,2}$ has a dummy value -1. All light curve parameters have been found unreliable ($R=1$). Also, because the model parameters have been found unreliable, no value has been computed for the time scale of change $T_{\rm C}$.

\begin{table}[t]
\caption{The format of the CPS output file for each individual datset.}
\begin{tabular}{l l l l}
\hline\hline
[SEG] = integer & $[t_0] =$ HJD & [SET] = integer & \\
$[t_1] =$ HJD & $[t_n] =$ HJD & $[\tau] =$ HJD & [IND] = integer \\
$[n] =$ integer & $[m_y] =$ mag & $[s_y] =$ mag & \\
$[K] =$ integer & $[\sigma_{\epsilon}] =$ mag & $[T_{\rm C}]$ = d & \\
$[M] =$ mag & $[\sigma_M] =$ mag & [R] = integer & \\
$[P] =$ d & $[\sigma_P] =$ d & [R] = integer & \\
$[A] =$ mag & $[\sigma_A] =$ mag & [R] = integer & \\
$[t_{\rm min,1}] =$ HJD & $[\sigma_{t_{\rm min,1}}] =$ d & [R] = integer & \\
$[t_{\rm min,2}] =$ HJD & $[\sigma_{t_{\rm min,2}}] =$ d & [R] = integer & \\
\hline
\end{tabular}
\label{coding}
\end{table}
\end{appendix}

\bibliographystyle{aa}
\bibliography{cps}

\end{document}